\documentclass[aps,prl,twocolumn,superscriptaddress, showpacs,floatfix]{revtex4-1}
\usepackage{graphicx}
\usepackage{dcolumn}
\usepackage{bm}
\usepackage{textcomp}
\begin{document}
\title{Microwave-driven Topological Resonant Excitations of Coupled Skyrmions}

\author{Han Wang}
\author{Yingying Dai}
\author{Teng Yang}
\author{Weijun Ren}
\author{Zhidong Zhang}
\email[E-mail: ]{zdzhang@imr.ac.cn}
\affiliation{Shenyang National Laboratory for Materials Science,
Institute of Metal Research and International Centre for Materials Physics,
Chinese Academy of Sciences, 72 Wenhua Road, Shenyang 110016,
PRC}
\date{\today}

\begin{abstract}
We study nonlinear dynamics of coupled Skyrmions and present a method for manipulating topological resonant excitations by a dual-frequency microwave field. Thiele's equation is extended by introducing a new effective mass associated with time derivative of topological density. Two coupled resonant modes endowed with the new effective mass are found in the coupled Skyrmions. Polygon-like resonant excitations are observed and modulated when the two modes are activated simultaneously by microwave field with commensurate frequency ratio. Quasiperiodic behavior of excitations are related to an incommensurate ratio. Numerical solutions based on the extended Thiele's equation for Skyrmion under a dual-frequency field agree well with the micromagnetic simulation results and clarify the importance of frequency value, frequency ratio of an external field and the effective mass to the polygon-like dynamics of Skyrmion. We also show the coupling between two Skyrmions can dominate their topological resonant excitations.
\end{abstract}
\pacs{
52.55.-s, 
75.78.Cd, 
12.39.Dc, 
75.50.Ss 
}

\maketitle
\section{Introduction}
Resonant excitations of topological objects have aroused a great deal of attention owe to both fundamental physics and potential application as microwave devices.\cite{ref30, ref33} Resonant excitations of vortices, which are always treated as massless particles with charges due to their small and local deformations,\cite{ref28, ref32, ref45} show only circular or elliptical trajectories in linear regime and stadium-like orbits in nonlinear regime.\cite{ref26, ref32} For dynamics of bubble, an inertial mass is considered due to the local deformation of one-dimensional (1D) domain wall.\cite{ref48}

Skrymion is another nontrivial topological object, which was recently observed experimentally in helical magnets,\cite{ref4, ref6, ref8, ref10,ref9,ref11,ref1} and also predicted to exist in coupled ordinary magnets.\cite{ref12, ref13} In contrast to vortices and bubbles, Skyrmions have the smooth global spin textures. Their deformations under resonant excitation will be nonlocal and so large that they can't be treated as particles or 1D domain walls, and the nonlinear dynamic equation has to include higher-order gyrotropic terms. The resonant excitation behavior are different from that of other topological objects with very local deformation.

Skyrmions are envisioned promising candidate for applications in spintronics, magnetic storage and microwave devices,\cite{ref15, ref16, ref23, ref2} because of their nanoscale size, high structural stability, low threshold current density to drive their motion, and intriguing magnetoelectric effect.\cite{ref6, ref17, ref18, ref2} Therefore, besides spin-polarized current,\cite{ref19, ref3} searching other methods of efficiently manipulating Skyrmions embedded in magnetic conductor and insulator is an important subject. The investigations of resonant excitation of Skyrmion or coupled Skyrmions have great significance to the Skyrmion dynamics and is crucial for their practical application.

In this work, we present an approach for manipulating topological resonant excitations of coupled Skyrmions by a dual-frequency microwave field. A new effective mass associated with topological density distribution is introduced into Thiele's equation to describe the dynamics of coupled Skyrmions with nonlocal deformation. We find two resonant modes with opposite sense of rotation under single-frequency microwave field. The two modes have dynamically coupled phase under a dual-frequency field, which has not been discovered before. By modulating the commensurate ratio of the dual-frequencies, we obtain polygon-like resonant excitation and are able to controllably change it. Quasiperiodic behavior is observed when the frequency ratio is incommensurate. Numerical solutions to the extended Thiele's equation are in best agreement with the micromagnetic simulation results and verifies that the effective mass is vital to the polygon-like resonant excitations of Skyrmion as well as the value of frequency and the frequency ratio of the external field. We also show the perpendicular coupling between two Skyrmions is crucial to their dynamics.

\section{Methods}
Resonance excitation of two coupled Skyrmions in Co/Ru/Co nanodisks under in-plane microwave magnetic fields was studied using {\footnotesize OOMMF} code.\cite{ref36}
The material parameters of hexagonal-close-packed (hcp) cobalt chosen include the saturation magnetization \emph{M$_s$}=1.4$\times$10$^6$ A/m, the exchange
stiffness \emph{A$_{ex}$}=3$\times$10$^{-11}$ J/m and the uniaxial anisotropy constant \emph{K$_u$}=5.2$\times$10$^5$ J/m$^3$ with the direction perpendicular to the nanodisk
plane. The interfacial coupling constant of the adjacent surfaces was -5$\times$10$^{-5}$ J/m$^2$ according to Ref. \onlinecite{ref37}. A dimensionless damping
$\alpha$ was 0.02. The radius \emph{R} of a Co/Ru/Co nanodisk was 100 nm. The thickness \emph{L} of Co was 18 nm, while that of Ru was 2 nm. The cell size was
2$\times$2$\times$2 nm$^3$, which is smaller than the exchange length of cobalt (about 4.94 nm). The in-plane microwave field was applied in x-direction to drive Skyrmion motion. The frequency of external field is equal or close to the eigenfrequency of system to make Skyrmion be resonant. The waveforms of microwave fields are $H sin(2 \pi ft)$ with
frequency of 1 GHz and 5 GHz and $H_{1}sin(2 \pi f_1t)+H_{2}sin(2 \pi f_2t)$ with $f_1/f_2$ of 1/2, 1/3, 1/4, 1/5, 1/6 and $H_{1}/H_{2}$ of 2/1, 3/1, 4/1, 5/1, 6/1, respectively.

\section{Effective mass}
To consider the nonlocal nature of large deformation, we assume local magnetization $\bm{S}(\bm{x},t)$ depends not only on Skyrmion core's position $\bm{X}(t)$ and velocity $\dot{\bm{X}}(t)$ but also on acceleration $\ddot{\bm{X}}(t)$: $\bm{S}(\bm{x}-\bm{X}(t),\dot{\bm{X}}(t),\ddot{\bm{X}}(t))$. We describe dynamics of Skyrmion through third-order dynamics equation of vortex:\cite{ref41, ref42}
\begin{eqnarray}
\bm{G_3}\times\dddot{\bm{X}}+\bm{F}'+\bm{G}\times\dot{\bm{X}}=\hat{\bm{e}}_i M_{ij}'\ddot{X}_j, \label{eq1} \\
M_{ij}'=-\gamma^{-1}S^{-2} \int d^2 x \bm{S}\cdot (\frac {\partial \bm{S}} {\partial x_i} \times \frac {\partial \bm{S}} {\partial \dot{X}_j}), \label{eq2}
\end{eqnarray}
where $\bm{G}$, $\bm{G}_3$, and $\bm{F}'$ are the gyrovector, third-order gyrovector, and the force acting on cores, respectively. $\hat{\bm{e}}_i$ is unit coordinate vector ($i (j) = x, y)$ and $\gamma$ the gyromagnetic ratio. Here, applying Wysin's transformation\cite{ref41} $\bm{R}(t)=\bm{X}(t) + \frac {M'} {|\bm{G}|^2} \bm{G}\times\dot{\bm{X}}(t)$, we can get:
\begin{eqnarray}
\bm{S}(\bm{x},t)=\bm{S}(\bm{x}-\bm{R}(t),\dot{\bm{R}}(t)),
\label{eq3}
\end{eqnarray}
where $\bm{R}(t)$ is topological charge center (guiding center) of Skyrmion.\cite{ref38}
Then we are able to extend Thiele's equation of the guiding center:
\begin{equation}
-\partial U/\partial\bm{R}+\mu \bm{H}+\bm{G}\times\dot{\bm{R}}=\hat{\bm{e}}_i M_{ij}\ddot{R}_j,
\label{eq4}
\end{equation}
where coefficient $\mu$ is a function of structural and magnetic parameters,\cite{ref45, ref46} $\bm{H}$ is an external magnetic field and \emph{M} is the new effective mass tensor with elements
\begin{eqnarray}
M_{ij}=-\gamma^{-1}S \int d^2 x \bm{n}\cdot (\frac {\partial \bm{n}} {\partial x_i} \times \frac {\partial \bm{n}} {\partial \dot{R}_j}), \label{eq5} \\
\dot{R}_j = \frac{\int x_j \dot{q} d^2 x} {\int q d^2 x}, \quad q \equiv \frac{1} {2} \epsilon_{i j} \bm{n}\cdot(\frac {\partial \bm{n}} {\partial x_i} \times \frac {\partial \bm{n}} {\partial x_j}),
\label{eq6}\
\end{eqnarray}
where $q$, $\epsilon_{i j}$, and $\bm{n}$ are topological density, the antisymmetric tensor, and the unit vector of local magnetization, respectively. The potential energy of the guiding centers:
\begin{equation}
U=\frac {1} {2}k\bm{R}_{t}^2+\frac {1} {2}k\bm{R}_{b}^2+U_{Sky-Sky}(d),
\label{eq7}
\end{equation}
where $k$, $d=|\bm{R}_t-\bm{R}_b|$, and $U_{Sky-Sky}(d)$ are stiffness coefficient, the distance of two guiding centers, and the magnetostatic coupling between two Skyrmions, respectively. The new effective mass is related to the time derivative of topological density $\dot{q}$, which makes it different from the mass $M_{ij}'$ of Eq. (\ref{eq2}). For the nonlocal deformation of topological object, the center of topological density and the mass associated with variation of topological density have to be used to comprehend its dynamical behavior.

\section{Results of Micromagnetic simulation}
\begin{figure}[b]
\includegraphics[width=0.5\columnwidth]{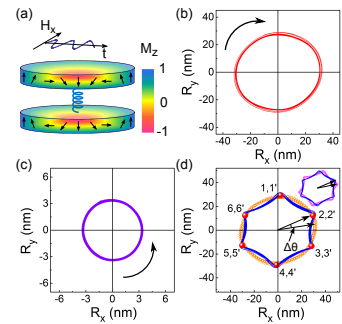}
\caption{(Color online) (a) The sketch of two coupled Skyrmions with opposite chiralities in a Co/Ru/Co nanodisk. (b), (c) and (d) are the trajectories of the Skyrmion on the top nanolayer under single frequency fields 1 GHz (b), 5 GHz (c), and dual-frequency field with the ratios $f_1/f_2$ of 1/5 and $H_{1}/H_{2}$ of 5/1 (d), respectively. The orange dot line and the pink dot line in the inset are the sum of orbits of 1 GHz and 5 GHz with amplitude ratio of 5/1 and 1/1, respectively. Curved arrows represent the sense of rotation. Numbers in (d) label the cusps of the hexagon during two periods.
\label{Fig1} }
\end{figure}
Resonant excitations of the two coupled Skyrmions are stimulated by a microwave magnetic field as shown in Fig. \ref{Fig1}. The field is 100 Oe of amplitude and applied along the +\emph{x} direction on the top nanolayer as shown in Fig. \ref{Fig1}(a). Trajectories of the guiding center for the top nanolayer under the field $Hsin(2 \pi ft)$ with frequencies of 1 GHz and 5 GHz are shown in Figs. \ref{Fig1}(b) and \ref{Fig1}(c), respectively. The frequencies of the external field are chosen to be the eigenfrequencies of a hexagonal trajectory for gyrotropic motion of coupled Skyrmions.\cite{ref12} Both resonant orbits are approximately circular, and radius of 30.5 nm in Fig. \ref{Fig1}(b) is much larger than that of 3.3 nm in Fig. \ref{Fig1}(c). We find that directions of their circulation are opposite, clockwise (CW) for the lower frequency mode while counterclockwise (CCW) for higher frequency mode. Compared with the resonant excitation of vortices,\cite{ref26, ref32} the rotation of Skyrmion depends not only on their polarities, but also strongly on the frequency of the microwave field due to their global spin textures. Mochizuki\cite {ref23} has studied the microwave-absorption spectra of a two-dimensional model, where a Skyrmion crystal is stabilized by the DMI. It turned out that for twofold spin-wave modes, with a small microwave magnetic field, the frequency difference of the two modes is small. However, we find that with the new effective mass, Eq. (\ref{eq5}), the two resonant excitation modes, far from being degenerate, exhibit a large frequency difference. When a dual-frequency microwave field $H_{1}sin(2 \pi f_1t)+H_{2}sin(2 \pi f_2t)$ with ${f_1}$ = 1 GHz and ${f_2}$ = 5 GHz is applied, the resonant trajectory of excitation is a hypocycloid, reminiscent of a hexagon instead of a circle as shown in Fig. \ref{Fig1}(d). Circular or elliptical, or even stadium-like orbits are common, while a hexagon is unusual in resonant excitation of vortices.\cite{ref46, ref26} Interestingly, comparing the trajectory of dual-frequency field with the sum trajectory of the two single frequency field, one can observe differences of amplitude and phase (about 12 degree), which indicates that two excitation modes is coupled with each other. The inset of Fig. \ref{Fig1}(d) shows the dynamical phase difference keep unchanged when the contribution of higher frequency mode was increased.
\begin{figure}
\includegraphics[width=0.9\columnwidth]{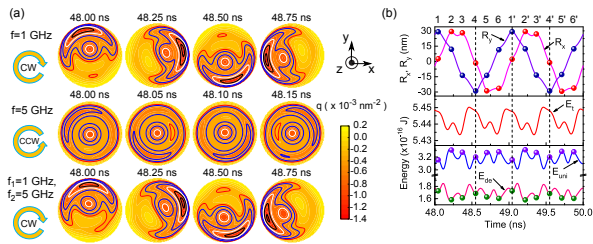}
\caption{(Color online) (a) The variation of topological density for the top nanolayer in one period under the different in-plane microwave magnetic fields. (b) The time dependence of $R_x$, $R_y$ and the total ($E_t$), uniaxial anisotropy ($E_{uni}$) and demagnetization ($E_{de}$) energies in two periods for the hexagonal steady orbit in Fig. \ref{Fig1}(d). Dots correspond to the cusps of the hexagon.
\label{Fig2} }
\end{figure}

To comprehend the striking features of hexagonal resonant excitation, Fig. \ref{Fig2}(a) shows time evolution of the topological density in one period for the steady orbits in Figs. \ref{Fig1}(b)-1(d). All the topological density distributions display a large global deformation, unlike very small and local deformations in vortex cores and in the narrow walls in bubbles.\cite{ref48, ref41} Clearly, this global deformation rotates like a rigid body, suggesting that the new effective mass associated with deformation of topological density shall be included to describe dynamics of coupled Skyrmions. The sense of rotation of the lower frequency mode is CW, while that in the higher one is CCW, which consists with the results in Figs. \ref{Fig1}(b)-\ref{Fig1}(d). When the dual-frequency microwave magnetic field is applied, the rotation sense is CW and the distribution of topological density is similar to that of the lower frequency mode with a slight difference, indicating that the hexagon is a superposition of CW and CCW modes with the CW mode dominating.

In order to elucidate the contribution of competition between different energies to stable hexagonal trajectory, Fig. \ref{Fig2}(b) shows time dependence of guiding center and the energies of Skyrmions. Dots correspond to the six cusps of the hexagon labeled in Fig. \ref{Fig1}(d). Obviously, $R_x$, $R_y$ and energies vary periodically with time and the periods are about 1 ns. Uniaxial anisotropy ($E_{uni}$) and demagnetization ($E_{de}$) energies take the most part of total energy. Interestingly, variation of $E_t$ is almost one order smaller than that of both $E_{uni}$ and $E_{de}$, suggesting competition between $E_{uni}$ and $E_{de}$ plays a substantial role in determining the stable hexagonal trajectory. At each cusp, $E_{de}$ locates valley and $E_{uni}$ locates peak, whereas $E_{de}$ reaches maximum and $E_{uni}$ reaches minimum between two cusps. The above analysis is also applicable to other types of polygonal trajectories.

\begin{figure}[t]
\includegraphics[width=0.5\columnwidth]{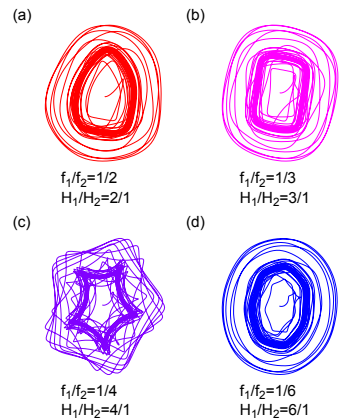}
\caption{(Color online) Transformation of topological resonant trajectory of the guiding center under different microwave magnetic fields $H_{1}sin(2 \pi f_1t)+H_{2}sin(2 \pi f_2t)$. (a), (b), (c), (d) $f_1$ is chosen to be 1.15 GHz and the ratio of $f_1/f_2$ is 1/2, 1/3 1/4 and 1/6, respectively. The corresponding ratio of $H_{1}/H_{2}$ is 2/1, 3/1, 4/1, and 6/1, respectively. Amplitude $H_{1}$ is 50 Oe.
\label{Fig3}}
\end{figure}

We have demonstrated that resonant excitation can be used to transform rotational trajectory of Skyrmion from approximately circle to hexagon by switching microwave field to dual frequencies from single one. Here we further show in Fig. \ref{Fig3} that tuning ratios in the dual-frequency microwave field is also beneficial for the controllability of Skyrmion dynamics. The frequency $f_1$ is chosen to be 1.15 GHz, which is the eigenfrequency of a circular trajectory for gyrotropic motion of coupled Skyrmions. Under microwave field with the eigenfrequency, dynamics of Skyrmion will exhibit a strong resonant excitation. Based on the strong resonant excitation, polygon-like trajectories can be obtained by tuning the frequency ratio of dual-frequency field. The frequency ratio $f_1/f_2$ is 1/2, 1/3, 1/4 and 1/6, whereas the amplitude ratio is 2/1, 3/1, 4/1, and 6/1 in Figs. \ref{Fig3}(a)-\ref{Fig3}(d), respectively. The steady orbit of the topological resonant excitation is indeed able to be transformed from hexagon to other polygons such as triangle, quadrangle, pentagon, and heptagon. Such controllability of resonant excitation under a microwave magnetic field would be useful to manipulation of Skyrmions and design of microwave devices.

\begin{figure}[b]
\includegraphics[width=0.5\columnwidth]{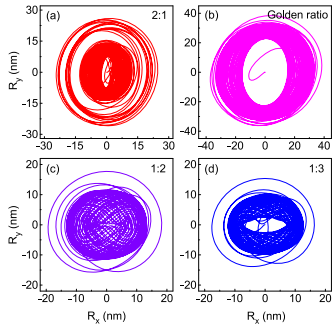}
\caption{(Color online) The quasiperiodic behavior of resonant excitation of coupled Skyrmion under microwave magnetic field with frequency $f_1$ = 1.15 GHz and $f_1/f_2$ = 1/golden ratio. (a), (b), (c), (d) The corresponding amplitude ratio $H_{1}/H_{2}$ is 2/1, golden ratio, 1/2, and 1/3, respectively. Amplitude $H_{1}$ is 50 Oe.
\label{Fig4} }
\end{figure}

In Fig. \ref{Fig4} the resonant excitation of coupled Skyrmion shows quasiperiodic behavior when the frequency ratio of microwave magnetic field is incommensurate and very close to the golden ratio. The trajectories of resonant excitations under single-frequency (rational number) microwave field are periodic circles. After introduced microwave with irrational number frequency, the system began to display quasiperiodicity. With the increase of amplitude ratio $H_{1}/H_{2}$ £¬the trajectories become more uncertain due to increasing effect of irrational number frequency. The nonlinear dynamical behavior was also found in a josephson-junction system under two ac sources, when the frequency ratio is golden mean.\cite{ref53}

\begin{figure}
\includegraphics[width=0.5\columnwidth]{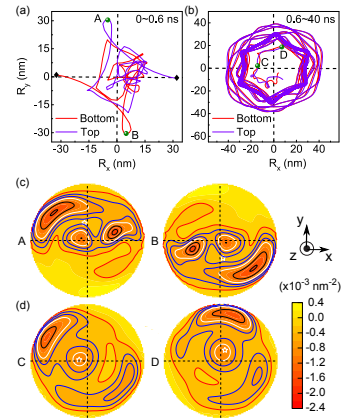}
\caption{(Color online) (a) and (b) Trajectories of the guiding centers of two Skyrmions in 0$\sim$0.6 ns and 0.6$\sim$40 ns, respectively, after replacing an applied constant magnetic field with the dual-frequency microwave field. Dark diamonds in (a) show the positions of two guiding centers at 0 ns. (c) and (d) The topological density distributions of points A (0.035 ns on top nanolayer), B (0.035 ns on bottom nanolayer), C (1.008 ns on top nanolayer) and D (1.216 ns on top nanolayer) labeled in (a) and (b), respectively. The white stars in (c) and (d) represent positions of the guiding center.
\label{Fig5} }
\end{figure}
We have to point out that coupling between two Skyrmions also plays a crucial role in the stability and controllability of coupled skyrmion dynamics, as illustrated in Eqs. (\ref{eq4}) and (\ref{eq7}) and discussed as follows. Figure \ref{Fig5} shows the influence of coupling on the trajectories and topological density distributions of Skyrmions. Firstly, two guiding centers are driven away by an external constant magnetic field along the +\emph{x} direction to $\pm$ 30 nm for top and bottom Skyrmions, respectively, as labeled by dark diamonds in Fig. \ref{Fig5}(a). The guiding centers of two Skyrmions move in opposite directions due to different chiralities and are finally decoupled. Then a dual-frequency microwave magnetic field ($f_1/f_2$=1/5 and $H_{1}/H_{2}$=5/1) replacing the constant field drives the two skyrmions to rotate. During 0$\sim$0.6 ns,
the two guiding centers of Skrymions keep at diagonal position from each other but with the same rotation senses of CCW. The trajectories are quadrangles other than hexagon, quite complex and unstable as shown in Fig. \ref{Fig5}(a). The topological density distributions of points A and B labeled in Fig. \ref{Fig5}(a) also show complicated patterns and large deformations in Fig. \ref{Fig5}(c). When getting close to each other at 0.6 ns due to a damping effect, the guiding centers of Skyrmions are coupled. Consequently, their rotation senses change from CCW to CW, and hexagonal trajectories revive, as shown in Fig. \ref{Fig5}(b). Compared with the topological density distributions at initial time, those of the coupled Skyrmions shown in Fig. \ref{Fig5}(d) start to take a shape similar to the stable ones previously shown in Fig. \ref{Fig2}(c), and rotate as a relative rigid body from point C to D (labeled in Fig. \ref{Fig5}(b)). The rotation sense changing indicates that Skyrmions dynamics depends not only on their polarities, but also on the eigenfrequencies of circulation, which is distinguished from vortex. The results also indicate the coupling between Skyrmions can be used to manipulate topological density distribution, thus having influence on effective mass and dynamics of the system.

\section{Numerical Solutions to the Extended Thiele's Equation}
\begin{figure}
\includegraphics[width=0.5\columnwidth]{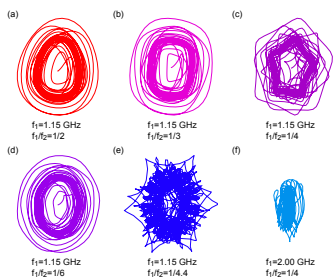}
\caption{(Color online) Numerical solutions to Eqs. (\ref{eq8}) and (\ref{eq9}) for Skyrmion driven by a dual-frequency field $H_x=H_1 sin(2\pi f_1t)+H_2sin(2\pi f_2t)$. The values of $f_1$, frequency ratio $f_1/f_2$ and the corresponding $H_1/H_2$ are (a) 1.15 GHz, 1/2 and 2/1, (b) 1.15 GHz, 1/3 and 3/1, (c) 1.15 GHz, 1/4 and 4/1, (d) 1.15 GHz, 1/6 and 6/1, (e) 1.15 GHz, 1/4.4 and 4.4/1, (f) 2.00 GHz, 1/4 and 4/1, respectively.
\label{Fig6} }
\end{figure}
To understand the polygon-like resonant excitations of coupled Skyrmions under a dual-frequency microwave field, numerical solutions to Eq. (\ref{eq4}) were obtained by the Runge-Kutta method for different frequency ratios, as shown in Fig. \ref{Fig6}. Because of the two Skyrmions moving almost synchronously, we can use an effective potential energy to replace the total potential energy of one Skyrmion. Equation (\ref{eq4}) with dissipation term considered for dynamics of Skyrmion in the top nanodisk can be rewritten as follows:
\begin{eqnarray}
\mu H_x-K R_x-G \dot{R_y}-D\dot{R_x} =M\ddot{R_x}, \label{eq8} \\
-K R_y+G \dot{R_x}-D\dot{R_y}=M\ddot{R_y}, \label{eq9}
\end{eqnarray}
where \emph{K} is the effective stiffness coefficient and \emph{D} the damping parameter.\cite{ref45} Here, for numerical solutions to Eqs. (\ref{eq8}) and (\ref{eq9}), $D=5.59\times10^{-14}$ J s/m$^2$, $G= 4\pi M_s L/\gamma=1.8\times10^{-12}$ J s/m$^2$ and $\mu=\pi \mu_0 R L M_s \xi=1.0\times10^{-14}$ kg m$^2$/(As$^2$) with $\xi\approx 0.93$ for Skyrmion in this work.\cite{ref45} Values of \emph{K} (0.013 J/m$^2$) and \emph{M} ($7.08\times10^{-23}$ kg) are calculated under zero field according to the eigenfrequencies of near 0.96 and 4.98 GHz for hexagonal trajectory of gyrotropic motion and 1.15 GHz for circular one.\cite{ref48, ref12}

When an external microwave magnetic field Hx is a single-frequency field, the resonance of system only displays circular or elliptical trajectory like resonance of magnetic vortex.
Polygon-like resonant excitations can be driven by a dual-frequency field $H_x=H_1 sin(2\pi f_1t)+H_2sin(2\pi f_2t)$ with $f_1$ of 1.15 GHz (one of the eigenfrequency of Skyrmion) and $f_2$ as an integral multiple of $f_1$, as shown in Figs. \ref{Fig6}(a)-\ref{Fig6}(d). The polygonal trajectories from numerical solution are in best agreement with those from micromagnetic simulations [Fig. \ref{Fig3}]. When $f_2$ is a non-integral multiple of $f_1$, a polygon-like orbit cannot be obtained.
 As an example, Fig. \ref{Fig6}(e) illustrates resonant excitation of Skyrmion under a field with $f_1/f_2$ of 1/4.4, which gives a different result from polygon-like resonance, indicating that the resonant dynamics of Skyrmion is sensitive to the frequency ratio of the dual-frequency field.
\begin{figure}
\includegraphics[width=0.4\columnwidth]{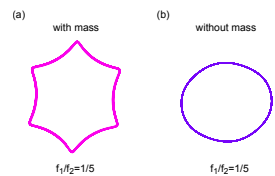}
\caption{(Color online) Numerical solutions to Eqs. (\ref{eq8}) and (\ref{eq9}) for Skyrmion driven by a dual-frequency field $H_x=H_1 sin(2\pi f_1t)+H_2sin(2\pi f_2t)$ with $f_1$ of 1.00 GHz (a) with and (b) without the effective mass term considered. The frequency ratio $f_1/f_2$ is 1/5, and the amplitude ratio $H_1/H_2$ is 5/1. Only steady trajectories are shown here.
\label{Fig7} }
\end{figure}

Actually, not only the frequency ratio, but also the value of frequency the value of frequency are of great importance to the dynamics of Skyrmion. Dynamics of skyrmions will show neither a strong resonance, nor polygon-like trajectory, if the frequency of the external field is not equal nor close to the eigenfrequency of Skyrmion. To confirm it, Fig. \ref{Fig6}(f) demonstrates the trajectory of Skyrmion under a field with frequency $f_1$ of 2.00 GHz, which is far away from the eigenfrequency of Skyrmion. In this case, though a dual-frequency field is applied, dynamics of Skyrmion does not show a polygon-like behavior.

To clarify the importance of the effective mass to the polygon-like dynamics of Skyrmion, Fig. \ref{Fig7} shows the numerical solutions to Eqs. (\ref{eq8}) and (\ref{eq9}) with and without the effective mass term. Parameters of Eqs. (\ref{eq8}) and (\ref{eq9}) in Fig. \ref{Fig7} are the same as used in Fig. \ref{Fig6}. An external field with $f_1$ of 1.00 GHz (one of the eigenfrequencies of coupled Skyrmions), $f_1/f_2$ of 1/5 and $H_1/H_2$ of 5/1 is applied on the system. Figure \ref{Fig7}(a) demonstrates the steady orbit of the numerical solution as the effective mass is considered. A hexagonal trajectory is obtained, which agrees well with the result of Fig. \ref{Fig1}(d) from the micromagnetic simulation. However, when the effective mass term is not considered, the result is totally different and an approximately circular trajectory obtained as shown in Fig. \ref{Fig7}(b). The results indicate that the effective mass is vital to the polygon-like trajectory.

From Figs. \ref{Fig6} and \ref{Fig7}, the factors governing the polygon-like excitations of Skyrmion can be listed as follows:
(1) The frequency $f_1$ of the external field has to be equal or close to the eigenfrequency of Skyrmion.
(2) A dual-frequency field with an integer frequency ratio $f_2/f_1$ is necessary.
(3) The effective mass of Skyrmion associated with time derivative of topological density is important to the polygon-like resonant excitations.

Note that twofold spin-wave modes excitation under a small microwave field are theoretically anticipated while only low-frequency mode has been experimentally observed. High-frequency mode may be mixed with conical mode or indistinguishable from breathing mode due to their similar resonance frequencies.\cite{ref23, ref2, ref51} Compared with the spin-wave modes, two modes of topological resonant excitations in present work have higher frequencies and larger frequency difference. Moreover, there is no existence of conical mode for Co/Ru/Co nanodisks. Therefore, we expect that the topological resonant modes are to be found in experiment and may contribute to the understanding of the collective excitation in Skyrmions.

\section{Conclusion}
In conclusion, we have numerically studied the topological resonant excitation of two coupled Skyrmions by micromagnetic simulations. Thiele's equation of guiding center incorporating a new effective mass term is used to comprehend two coupled resonant modes of Skyrmions with nonlocal deformation. Rotational trajectories have been controllably transformed from circular to different polygonal shapes by tuning commensurate frequency ratio of a dual-frequency microwave field with $f_1$ chosen to be the eigenfrequency of Skyrmion. Quasiperiodic behavior of Skyrmion dynamics is observed when the frequency ratio is incommensurate. Numerical solution to the extended Thiele's equation clarifies the importance of the value of $f_1$, the frequency ratio $f_1/f_2$ of a dual-frequency field and the effective mass to the polygon-like resonant excitations of Skyrmion. We have also shown the coupling between two Skyrmions has an obvious effect on their dynamics. These findings contribute to a better understanding of Skyrmion dynamics and may open a new avenue to the manipulation of spin textures by a dual-frequency microwave field.

\begin{acknowledgments}
We appreciate helpful discussions with Prof. Peter F. de Chatel. The work is supported by the National Basic Research Program (No. 2010CB934603) of China, Ministry of Science and Technology China and the National Natural Science Foundation of China under Grant No. 51331006.
\end{acknowledgments}


\end{document}